\documentclass[iop,apjl]{emulateapj}

\usepackage{natbib}
\slugcomment{The Astrophysical Journal}
\shorttitle{The infrared colors of the Sun}
\shortauthors{Casagrande et al.}

\newcommand{\teff}{T_{\rm{eff}}}
\newcommand{\logg}{\log g}
\newcommand{\feh}{\rm{[Fe/H]}}
\newcommand{\mic}{\mu\rm{m}}
\newcommand{\ubvri}{$UBV(RI)_\mathrm{C}\,$}

\begin{document}

\title{THE INFRARED COLORS OF THE SUN}

\author{
        L.\,Casagrande   \altaffilmark{1},
        I.\,Ram\'irez    \altaffilmark{2},
        J.\,Mel\'endez   \altaffilmark{3},
        M.\,Asplund      \altaffilmark{1},
      }

\altaffiltext{1}{Research School of Astronomy \& Astrophysics, Mount Stromlo 
                 Observatory, The Australian National University, ACT 2611, 
                 Australia}
\altaffiltext{2}{McDonald Observatory and Department of Astronomy,
                 University of Texas at Austin, 1 University Station, C1400
                 Austin, Texas 78712-0259, USA}
\altaffiltext{3}{Departamento de Astronomia do IAG/USP, 
                 Universidade de S\~ao Paulo,
                 Rua do M\~atao 1226, S\~ao Paulo, 05508-900, SP, Brasil}

\email{luca@mso.anu.edu.au}

\begin{abstract}

Solar infrared colors provide powerful constraints on the stellar effective 
temperature scale, but to this purpose they must be measured with both 
accuracy and precision. 
We achieve this requirement by using line-depth ratios to derive in a model 
independent way the infrared colors of the Sun, and use the latter to test the 
zero-point of the \cite{c10} effective temperature scale, confirming its 
accuracy.
Solar colors in the widely used 2MASS $JHK_s$ and WISE $W1\,W2\,W3\,W4$ 
systems are provided: $(V-J)_\odot = 1.198$, $(V-H)_\odot=1.484$, 
$(V-K_s)_\odot=1.560$, $(J-H)_\odot = 0.286$, $(J-K_s)_\odot = 0.362$, 
$(H-K_s)_\odot = 0.076$, $(V-W1)_\odot = 1.608$, $(V-W2)_\odot = 1.563$, 
$(V-W3)_\odot = 1.552$, $(V-W4)_\odot = 1.604$. A cross check of the effective 
temperatures derived implementing 2MASS or WISE magnitudes in the infrared flux 
method (IRFM) confirms that the absolute calibration of the two systems agree 
within the errors, possibly suggesting a 1\% offset between the two, thus 
validating extant near and mid infrared absolute calibrations. While 
2MASS magnitudes are usually well suited to derive $\teff$, we find that a 
number of bright, solar-like stars exhibit anomalous WISE colors. In most 
cases this 
effect is spurious and traceable to lower quality measurements, although for a 
couple of objects ($3\pm2$\% of the total sample) it might be real and hints 
towards the presence of warm/hot debris disks.

\end{abstract}


\section{INTRODUCTION}

Photometric systems and filters carry information on various fundamental 
stellar properties, such as effective temperature ($\teff$), metallicity 
($\feh$) and surface gravity ($\logg$). Also when studying more 
complex systems, integrated magnitudes and colors of stars can be used to 
infer properties of the underlying stellar populations, by interpreting 
observations via theoretical population synthesis models. 
However, stars with well known physical parameters and colors are needed to 
establish how observed photometric data must be translated into physical 
quantities and placed on an absolute scale. The absolute calibration of 
photometric systems precisely deal with this matter and it has a long and 
noble history, especially in using solar-type stars to this purpose 
\citep[e.g.][]{j65,w81,c85,r08}.
Arguably, the star with best known parameters as well as the most important 
benchmark in astrophysics is the Sun, but for obvious reasons it can not be 
observed with the same instruments and under the same conditions applied to 
distant stars, thus making virtually impossible to directly measure its 
colors \citep{sk57}.

Photometry of stars with stellar properties very similar to the Sun 
provides a way to cope with this limit, although it is not obvious how to 
identify stars satisfying such a condition in first place. Linking photometric 
measurements to stellar parameters is in fact the goal, and selecting Sun-like 
stars based on colors would clearly introduce a circular argument.
On the other hand, spectroscopy provides an excellent way of determining 
$\teff$, $\logg$ and $\feh$ in stars, and it is routinely used to this 
purpose, although it can be heavily model-dependent. 
Nevertheless, this major limit is easily overcome when restricting to a purely 
differential analysis of stars 
with spectra largely identical to a reference one. If the latter is solar, it 
is thus possible to identify the stars most closely resembling the Sun, the so 
called solar-twins\footnote{According to their increasingly similarity to the 
Sun, stars can be classified as solar-like, solar-analogs and solar-twins 
\citep{cds86}. The term ``solar-twin'' does not imply that the stars were born 
together with the Sun.}.

Over the last few years, some of us \citep{m06,m07,m06,m09,r09} have conducted 
a systematic search aimed to characterise and discover the best solar-twins in 
the local $\sim 100$~pc volume, starting from an initial 
sample of about one hundred stars in the Hipparcos catalogue chosen to be 
broadly consistent with being solar-like. For each candidate, high resolution, 
high signal-to-noise observations were conducted and compared to solar 
reference spectra (which in fact are reflected Sun-light of asteroids) obtained 
with the same instrumentation and within each observing runs \citep[at 
McDonald and Las Campanas observatories, see Section 2 in][]{r12}.

Because the procedure adopted to identify solar-twins does not assume any a 
priori $\teff$, solar-twins have already been used to set the zero-point of the 
effective temperature scale via the infrared flux method \citep[IRFM,][]{c10}. 
The effective temperature scale is then a basic ingredient for measuring 
metallicities and, by comparison with theoretical isochrones, to derive 
stellar ages. 
Thus, the zero-point of the effective temperature scale directly impacts basic 
constraints of Galactic chemical evolution models (e.g~the metallicity 
distribution function and the age-metallicity relation) as well it is important 
to correctly interpret the Sun in a Galactic context 
\citep[e.g.][]{n04,c11,dat12}. Because of its far reaching implications, we 
have continued to investigate this topic \citep[see also][for a comparison 
between the angular diameters measured by interferometry with those obtained 
via the IRFM]{huber}; in particular we have conducted dedicated observations to 
overcome the major bottleneck in linking stellar parameters to photometry, 
i.e.~the availability of homogeneous and high accuracy photometric data.
In \cite{m10} we have presented new Str\"omgren $uvby$ observations of 
more than seventy solar-analogs and derived the colors of the Sun in this 
system, which then have been used to investigate the zero-point of various 
metallicity scales. Similarly, in \cite{r12} we have presented new \ubvri 
photometry of 80 solar-analogs and derived solar colors in the widely used 
Johnson-Cousins system, obtaining a definitive value for the long debated 
value of $(B-V)_\odot = 0.653 \pm 0.003$. 

In this paper we finally focus on the infrared colors of the Sun and the 
tight contraints they can provide on the $\teff$ scale. In fact, even though 
it is possible to use Str\"omgren and Johnson-Cousins colors \citep{m10,r12}, 
infrared ones are better suited to this purpose, being nearly 
independent on blanketing and surface gravity effects for the spectral types 
considered here \citep[e.g.][]{be98}. 

In addition to this motivation, highly standardised and precise infrared 
photometry is nowadays available from all-sky surveys, essentially defining 
new standard systems for the years to come: 2MASS in the near-infrared, and 
the WISE satellite in the mid-infrared. Accurate solar colors in these two 
systems are thus crucial for a numbers of purposes. Most importantly, the 
reliability at which infrared measurements can be used to infer stellar 
properties must also be assessed: while 2MASS data can be confidently adopted 
in most cases, a number of stars seem to exhibit anomalous WISE colors. We 
find that most of those are artifacts which can be avoided by imposing more 
stringent observational constraints, although in a few cases they might be 
real and indicate the presence of debris disks.

\section{SAMPLE AND PHOTOMETRIC DATA} \label{s:sample}

Our sample consists of the 112 stars used in \cite{r12}, from whom we have 
adopted 
$V$ magnitudes and stellar parameters. The latter were derived using 
excitation and ionization equilibrium conditions \citep[][and references 
therein]{r12}, and because of the strictly differential analysis with respect 
to the solar reference spectrum, the impact of systematic errors is minimised 
across the limited parameter space covered by our sample. Although a few stars 
in our sample might have faint companions, the effect seems negligible for 
2MASS, but we detected a few anomalous stars in the WISE colors. As we discuss 
later, these stars were discarded from the analysis.

For each star we have queried the 2MASS $J\,(1.25\mic)$ $H\,(1.65\mic)$ 
$K_s\,(2.17\mic)\,$ and the WISE 
$W1\,(3.4\mic)\,W2\,(4.6\mic)\,W3\,(12\mic)\,W4\,(22\mic)$ catalogues 
\citep[][respectively]{cutri03,wise} for photometry. 
Some of the bright targets have saturated or unreliable 2MASS magnitudes; to 
retain the best data, in a given band we consider only observations having 
photometric quality flag ``A'', read flag ``1'', blend flag ``1'' (i.e.~one 
component fit to the source), contamination and confusion flag ``0'' 
(i.e.~source unaffected by known artifacts)\footnote{See http://www.ipac.caltech.edu/2mass/releases/allsky/doc/sec2$\_$2a.html}. This set of flags 
automatically retains stars with photometric uncertainty (``msigcom'') in a 
given band better than $0.06$~mag, and for the full sample mean errors are 
$0.02$~mag in $J$ and $K_s$ and $0.03$~mag in $H$ band.

\begin{figure*}
\includegraphics[scale=0.6]{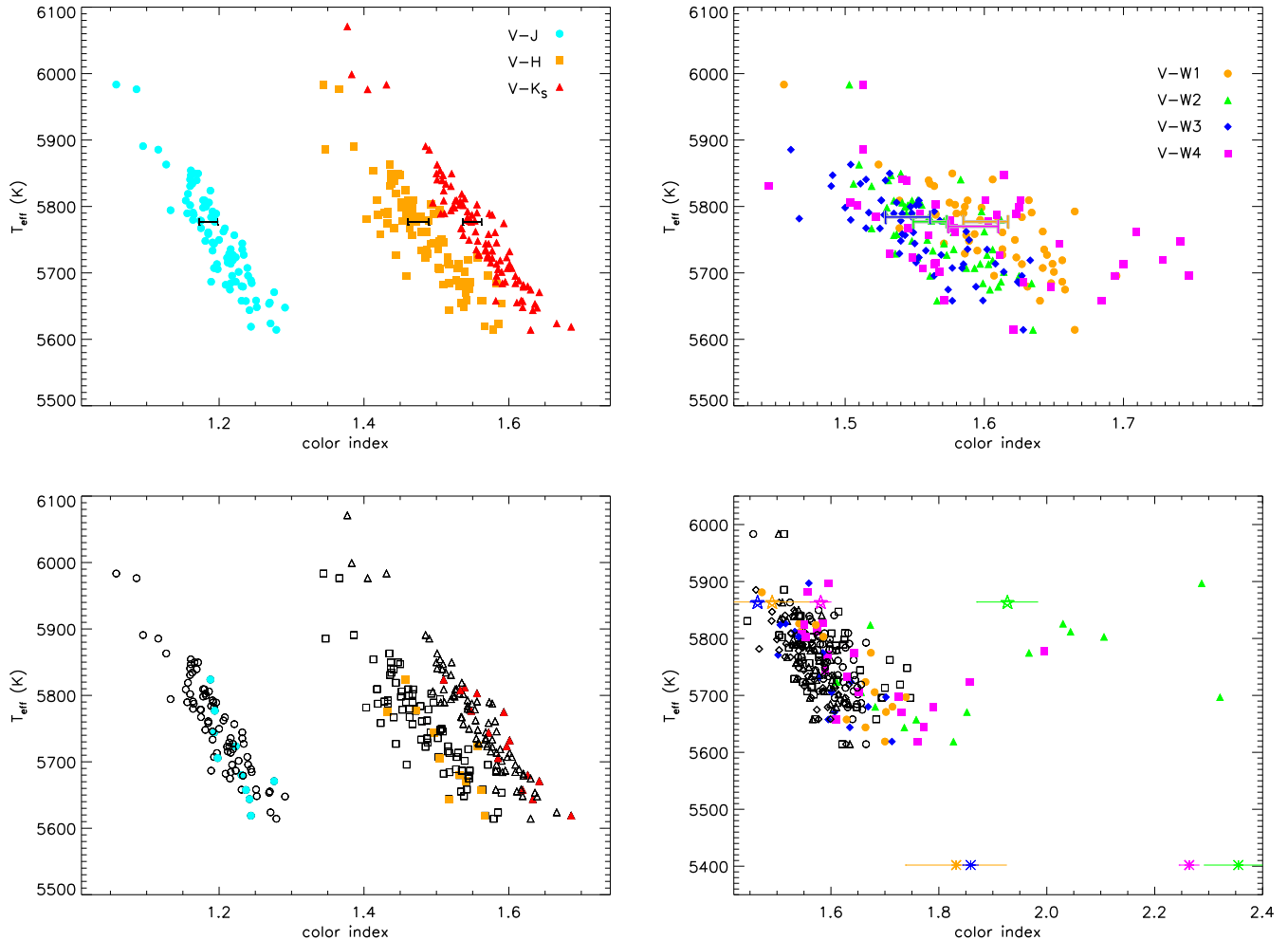}
\caption{Color indices versus effective temperatures obtained from the IRFM. 
  Top panels: using only stars satisfying the photometric criteria discussed 
  in Section \ref{s:sample}. The colors of the Sun are also shown with the 
  uncertainties derived via the IRFM (Table \ref{t:2}; notice that in the 
  right panel $T_{\rm{eff,\odot}}$ is offset by a few K for clarity purposes). 
  Bottom panels: objects having $V-W2 \ge 1.65$ or $V-W4 \ge 1.80$ but 
  satisfying all other photometric and quality constraints of Section 
  \ref{s:sample} are highlighted in colors. Stars from the upper panels are now 
  shown with empty circles. In the bottom right panel the effective 
  temperature and color ranges have been increased to include HD~69830 
  (asterisk); also shown is HD~72905 (pentagram). For both objects WISE's 
  photometric error bars are shown (see further discussion in Section 
  \ref{s:disc}).}
\label{f:fit}
\end{figure*}

Similarly, for WISE observations we restrict our analysis to measurements 
consistent with being point sources ($\rm{ext} = 0$, meaning that no band has 
a reduced $\chi^2>3$ and the source is not within 5 arcsec of a 2MASS Extended 
Source Catalogue entry), unaffected by known artifacts ($\rm{ccf} = 0$), 
quality flag ``A'' and variability flag ``n'' or $0-5$ (i.e.~most likely not 
variable)\footnote{http://wise2.ipac.caltech.edu/docs/release/allsky/expsup/sec2$\_$2a.html}. The ``A'' quality flag implies a signal-to-noise ratio higher 
than $10$, automatically curbing large photometric uncertainties. 
As detailed in the WISE Explanatory Supplement\footnote{http://wise2.ipac.caltech.edu/docs/release/allsky/expsup/sec6$\_$3d.html}, the $W1-4$ channels 
saturate at $\sim8.1,6.7,3.8,-0.4$~mag respectively, although fits to the 
unsaturated wings of the PSF allow viable magnitudes to be obtained up to 
$2.0,1.5,-3.0$ and $-4.0$~mag. Given the brightness of our targets, this is 
never a concern as the saturated pixel fraction is on average $0.07$ in $W1$ 
and it essentially drops to zero in the other bands.
Finally, to 
further decrease the possibility of having spurious identifications we also 
require each WISE source to have a 2MASS point source counterpart associated 
with it. All WISE sources identified with the above constraints have a 
position offset smaller than 3 arcsec (average $1.5$) with respect to the 
target coordinate. 
Some $20-30$ percent of the stars do fall severely apart from the main locus 
of the color-$\teff$ 
relations, especially in $W2$ and $W4$ band (no such effect is visible using 
2MASS), although all quality records listed above are fulfilled 
(Fig.~\ref{f:fit});
somewhat arbitrarily we exclude stars having $V-W2 \ge 1.65$ and $V-W4
\ge 1.80$ and we shall briefly discuss this in Section \ref{s:disc}. 
Altogether, this set of choices limits the mean (max) error to $0.03$ ($0.05$), 
$0.02$ ($0.02$), $0.02$ ($0.02$), $0.07$ ($0.11$) in $W1, W2, W3, W4$ 
respectively. 

\section{THE SOLAR INFRARED COLORS}\label{s:main}

For solar-type stars, $V-J$, $V-H$ and $V-K_s$ are known to display a 
remarkably tight correlation with $\teff$, while being nearly independent of 
other parameters such as $\feh$ and $\logg$. The strong temperature sensitivity 
in solar-type stars is due to the long wavelength baseline, 
which almost brackets the region of maximum flux in these stars, covering part 
of the spectrum with similar continuum opacity but differing temperature 
sensitivity to the Planck function; this argument continues to hold also when 
replacing 2MASS with WISE filters (Fig.~\ref{f:fit}).

In the literature there are various calibrations relating optical/near-infrared 
indices to effective temperatures of giants and dwarfs 
\citep[e.g.][]{r80,al96,rm05,c10,bo12a}; for the set of filters used in this 
work, it can be easily estimated that a change of about $0.03$~mag in solar 
colors implies an uncertainty of about $50$~K on the zero point of the 
$\teff$ scale. Therefore, to check the reliability of various $\teff$ scales, 
accurate and precise colors must be derived. This is done model independently
in Section \ref{s:method1}, while in Section \ref{s:method2} we check upon the 
zero-point of our effective temperature scale.

For this reason, it is important that the colors under investigation are 
obtained directly, without 
resorting on transformations between different systems. \cite{co87a,co87b} 
provides relations between the Str\"omgren and Johnson-Cousins photometry, and 
more recently, \cite{b8,b11} have derived an extensive set of color 
transformations relating the 2MASS, WISE and $BV(RI)_\mathrm{C}$ systems. Using 
those, the 
solar colors derived in \cite{m10,r12} and here are usually reproduced within 
$0.05$~mag (with better performances when transforming from the Str\"omgren 
system), although certain color combinations are offset by as much 
$0.10$~mag, nevertheless still consistent with the standard deviation of the 
transformations reported in \cite{b8,b11}.

\subsection{Spectral Line-Depth Ratios} \label{s:method1}

\begin{figure}
\includegraphics[scale=0.7]{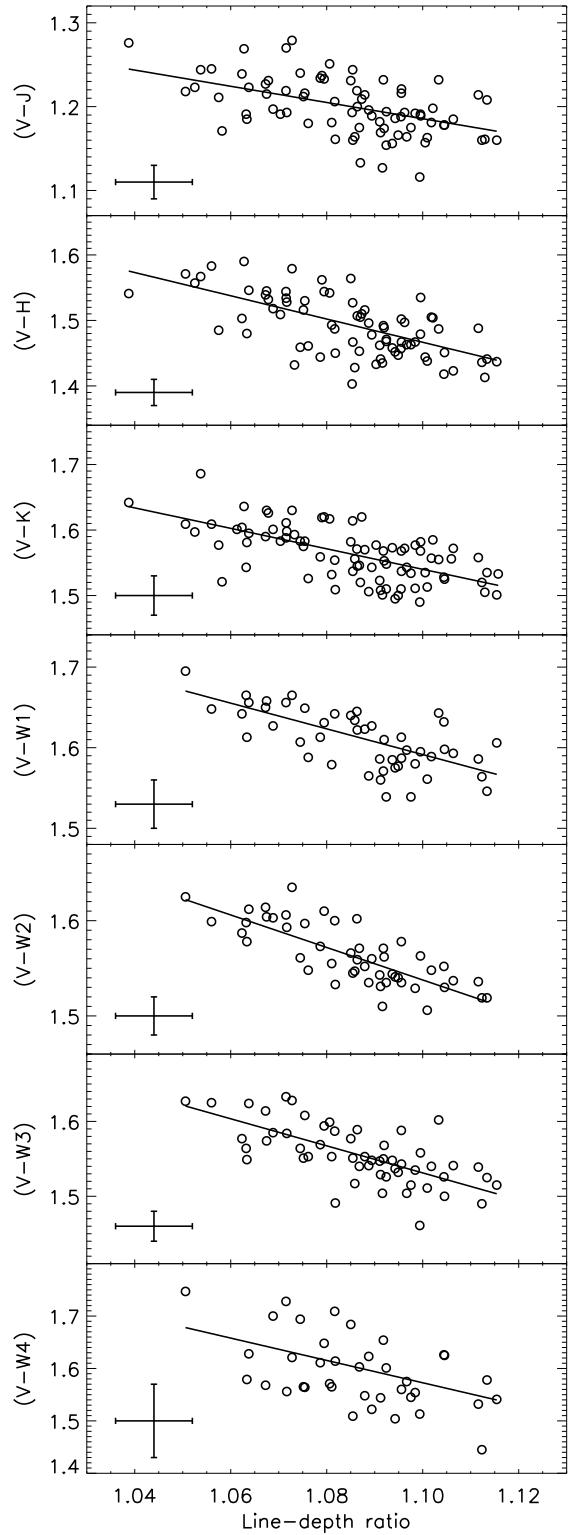}
\caption{Observed colors as a function of line-depth ratio for the 6119.53 
  \AA\ (V I), 6145.02 \AA\ (Si I) pair. The solid line is a linear fit to the 
  data.}
\label{f:ldr}
\end{figure}

Here we use the line-depth ratios (LDRs) technique as described in \cite{r12} 
to derive the infrared colors of the Sun in a model independent way. Briefly, 
this technique exploits the fact that the ratios of depths of spectral line 
pairs with very different excitation potential are excellent $\teff$ indicators 
--thus correlating well with observed colors-- essentially independent of 
$\feh$ and $\logg$ \citep[e.g.][]{g94}. For main sequence stars having 
$v\,\sin\,i \lesssim 5\,\rm{km\,s}^{-1}$, as the case of our stars, LDRs are 
also weakly depended on rotational broadening \citep{bia07}. For each set of 
line pairs \citep[from][]{kov} we measured the ratios in all stars of our 
sample and linearly fitted those ratios as function of the color index under 
consideration (Fig.~\ref{f:ldr}), after the exclusion of stars not satisfying 
the photometric quality requirements discussed in Section \ref{s:sample}. 
From each fit the standard deviation of the fit minus data residual
($\sigma_{\rm{fit}}$) was also obtained.
Notice that only line pairs for which the color vs.~LDR slope was 
greater than $0.3$ have been used. Slopes shallower than this imply a lower 
sensitivity, leading to larger errors in the derived solar color. Since the 
slope errors are about $0.03$, this criterion is equivalent to a $10\,\sigma$ 
cut. An example of all line pairs used to derive the solar $(V-K_s)$ color is 
given in Table \ref{t:ldr}. The interpolation of those fits at the solar ratio 
(measured in the reflected Sun-light of asteroids with the same procedure used 
for stars) returns the color index of the Sun. Since we have nine 
reflected Sun-light observations, nine solar line-depth ratio values are 
available for each line pair, resulting in nine solar colors. Columns 7 and 8 
of Table \ref{t:ldr} provide the mean and standard deviation 
($\sigma_{\rm{ss}}^2$) of those nine values.
For each color index there are usually about one hundred pairs available, 
thus making possible to derive extremely robust colors (Table \ref{t:1}), 
using the weighted mean of the values obtained from each line-depth ratio, 
where the weight $w$ is $1/w=\sigma_{\rm{fit}}^2+\sigma_{\rm{ss}}^2$ 
\citep[see also][]{r12}.

\setcounter{table}{1}
\begin{deluxetable}{ccc}
\tablewidth{0pc}
\tabletypesize{\small}
\tablecaption{Solar colors inferred from LDR Measurements.}
\tablehead{\colhead{color} & \colhead{value} & \colhead{$N_\mathrm{pairs}$}}
\startdata
 $(V-J)$   & $1.198\pm0.005$ &  87 \\
 $(V-H)$   & $1.484\pm0.009$ & 102 \\
 $(V-K_s)$ & $1.560\pm0.008$ & 100 \\
$(V-W1)$   & $1.608\pm0.008$ & 101 \\
$(V-W2)$   & $1.563\pm0.008$ & 102 \\
$(V-W3)$   & $1.552\pm0.009$ & 103 \\
$(V-W4)$   & $1.604\pm0.011$ &  92 \\
\enddata
\label{t:1}
\end{deluxetable}

\subsection{Color-$\teff$ Relations} \label{s:method2}

\begin{deluxetable*}{ccccc} 
\tablewidth{0pc}
\tabletypesize{\footnotesize}
\tablecaption{Solar colors inferred from different $\teff$ measurements: 
spectroscopic (i.e.~excitation and ionization equilibrium) and the IRFM 
implemented using Tycho2$+$2MASS 
(TY2M) or Johnson-Cousins$+$2MASS (JC2M) photometry.}
\tablehead{\colhead{color} & \colhead{spectroscopic} & \colhead{$\rm{IRFM_{TY2M}}$} & \colhead{$\rm{IRFM_{JC2M}}$} & \colhead{$N_\mathrm{stars}$}}
\startdata
 $(V-J)$   & $1.207\pm 0.013$ & $1.197\pm 0.013$ & $1.185\pm 0.013$ & 87 \\
 $(V-H)$   & $1.499\pm 0.014$ & $1.489\pm 0.013$ & $1.475\pm 0.014$ & 87 \\
 $(V-K_s)$ & $1.572\pm 0.013$ & $1.563\pm 0.013$ & $1.549\pm 0.013$ & 95 \\
 $(V-W1)$  & $1.620\pm 0.015$ & $1.613\pm 0.014$ & $1.601\pm 0.016$ & 52 \\
 $(V-W2)$  & $1.576\pm 0.011$ & $1.570\pm 0.011$ & $1.561\pm 0.012$ & 56 \\
 $(V-W3)$  & $1.564\pm 0.013$ & $1.558\pm 0.014$ & $1.545\pm 0.016$ & 59 \\
 $(V-W4)$  & $1.610\pm 0.021$ & $1.601\pm 0.018$ & $1.592\pm 0.018$ & 42 \\
\enddata
\label{t:2}
\end{deluxetable*}

The technique presented in Section \ref{s:method1} provides an elegant and 
model independent way of determining the colors of the Sun. Alternatively, it 
is possible to perform a multiple regression of the stellar parameters 
(e.g.~$\teff, \logg, \feh$) relevant to a given color index and derive that of 
the Sun by solving with respect to its parameters \citep[e.g.][]{h06,m10,r12}. 
Works focusing on the $\teff$ scale adopt essentially the same approach, where 
the colors of the Sun are inferred by reverting polynomial 
$\rm{color}-\feh-\teff$ relations derived for dwarf stars 
\citep[e.g.][]{rm05,c06}. While indices in Table \ref{t:1} tightly correlate 
with $\teff$, because of the relatively narrow parameter space covered by of 
our stars, we have verified that they do not display any significant 
dependence on $\logg$ nor $\feh$. In fact, performing a multiple linear 
regression with respect to all three parameters or $\teff$ only did not 
improve upon the residual nor changed within $0.001$ mag the values derived. 
Such a simple linear relation has also the advantage of making straightforward 
the connection between a given color index and the underlying $\teff$ scale. 
Depending on the band considered, several tens of stars survive the quality 
cuts we impose on 2MASS/WISE photometry (Section \ref{s:sample} and Table 
\ref{t:2}). Using the spectroscopic (excitation and ionization equilibrium) 
temperatures determined for the full sample of solar-analogs returns solar 
indices 
systematically redder by $\sim 0.01$~mag with respect to those obtained via 
LDRs. This implies that on average our spectroscopic $\teff$ are overestimated 
by about $20$~K, in agreement with what found by \cite{r12} using optical 
indices. 

For all stars in our sample we also run the IRFM to derive effective 
temperatures uncorrelated to the spectroscopic analysis. In fact, the IRFM is 
essentially model independent and very little affected by the metallicity and 
surface gravity of each star, the most relevant ingredient being the absolute 
calibration of the photometric systems adopted. Since all these stars are 
nearby, reddening is also not a concern as confirmed by using intrinsic 
Str\"omgren color calibrations \citep{m10}. 
In \cite{c10} the zero point of the IRFM scale was calibrated using 
solar-twins having Tycho2 and 2MASS photometry. Now, the availability of 
\ubvri magnitudes allows to check whether this is also the case when using 
instead the Johnson-Cousins system\footnote{For an implementation of the IRFM 
using WISE magnitudes instead of 2MASS see Appendix \ref{s:app}. Here we 
prefer to use the 2MASS system because of the better quality data.}. Replacing 
the Tycho2 with the Johnson-Cousins system in the IRFM returns effective 
temperatures cooler by $26 \pm 4$~K ($\sigma = 41$~K) which is within the 
zero-point uncertainty of our effective temperature scale. The same conclusion 
is obtained restricting the analysis to solar-twins only\footnote{I.e. stars 
having spectroscopic stellar parameters within $1.4\sigma$ from the solar 
ones, in accordance with the criterium used by \cite{r12}.} for which we 
obtain a median/mean $\teff$ of $5787/5786$~K and $5762/5750$~K using Tycho2 
and Johnson-Cousins photometry, respectively. These differences are mirrored in 
the color indices of Table \ref{t:2}: those inferred using Johnson-Cousins in 
the IRFM are in fact slightly bluer than obtained via LDRs, by an amount which 
would be almost perfectly offset should the effective temperatures increase by 
$20$~K. As expected, $(V-J)_\odot$, $(V-H)_\odot$ and $(V-K_s)_\odot$ derived 
here agree extremely well with those obtained by \cite{c10} reverting 
$\rm{color}-\feh-\teff$ polynomials defined over a much wider parameter 
space\footnote{Incidentally, using optical and infrared solar colors from LDRs 
in the aforementioned polynomials returns an average $\teff=5755\pm22$~K.}.

Effective temperatures determined using Tycho2 photometry in the IRFM return 
color indices which agree almost perfectly with LDRs. These results confirm 
the overall good consistency obtained implementing different photometric 
systems in the IRFM, with systematic uncertainties at the level of about 
$20$~K, i.e. $\sim 0.01$~mag in colors. Errors in Table \ref{t:2} take this 
into account, by adding such zero-point systematic to the uncertainties 
derived analytically from the fits. 

\section{The WISE through excess of infrared is made a fool?}\label{s:disc}

As discussed in Section \ref{s:sample}, stars having exceedingly red indices 
in WISE (Table \ref{t:anom}) were not used to derive solar colors, although 
all other photometric quality constraints were satisfied (apart from a few 
cases having in a given band $\rm{ccf} \ne 0$ or signal-to-noise lower than 
10, bands which were always excluded from the analysis). The unusual colors of 
these stars are clearly visible in the bottom right panel of Figure 
\ref{f:fit}. The bands most strikingly affected are $W2$ and partly $W4$, while 
$W1$ and $W3$ seem only slightly offset to the red with respect to the main 
locus defined by the full sample. This sort of signature ($W1-W2\gtrsim 0.3$) 
would not be entirely unexpected if looking at brown dwarfs. In fact, WISE's 
two shortest bands are designed to optimize sensitivity to this class of 
objects, by probing their deep $\rm{CH}_4$ absorption band at $\sim 3.3\mic$ 
($W1$)  and the region relatively free of opacity at $\sim4.6\mic$ ($W2$) 
where their Planck function approximately peaks \citep[e.g.][]{wwise,ma11,ki11}.

To quantify the amount of contamination expected from a potential low mass star 
companion we combine a synthetic MARCS \citep{g08} solar spectrum with that 
of a late type 
M dwarf ($\teff=2500$~K) of the same metallicity, and assuming a secondary to 
primary radius ratio of $0.2$ \citep{bo12b}, we conclude that the effect on the 
color indices shown in Figure \ref{f:fit} would be of order $0.01$ mag. 
Therefore, even if the spectral features of a brown dwarf could account for the 
red $W1-W2$ index we observe, the overwhelming flux of the primary makes this 
solution not viable even for an M dwarf, in addition to the fact that it would 
still be difficult not to affect $W3$ . Neither the alignment/confusion with 
extragalactic sources (which would be considerably fainter than our objects, 
see below) or cool (sub)-stellar objects is likely. The angular resolution of 
WISE passes from $6.1$~arcsec in $W1$ to $12$~arcsec in $W4$; using the higher 
resolution of 2MASS, none of the targets discussed here has more than one 
counterpart within $12$~arcsec. All anomalous sources have $W1$ brighter than 
$7.6$~mag ($5.9$~mag if considering the reddest $V-W2 > 1.8$). 
Using this constraint with the previous synthetic model ($\teff=2500$~K) would 
imply the presence of cool M dwarfs closer than $7.5$ ($3.5$) pc and brighter 
than $V \sim 17$ ($15.5$)~mag. Using instead a synthetic brown dwarf spectrum 
\citep[$\teff=1000$~K, ][]{bur06} and adopting $R=0.1R_{\odot}$ the above 
estimates would change into distances closer than $1$ ($0.5$)~pc and $V$ 
magnitudes brighter than $\sim21$ ($19.5$)~mag, thus making extremely unlikely 
the superposition of brown dwarfs to our solar-like stars. 

Interpreting the color anomaly as mid-infrared emission could hint toward the 
presence of warm/hot debris disks, even though this class of object is thought 
to be very rare compared to most of the known cold Kuiper-belt type disks, 
especially around old main sequence stars \citep[e.g.][]{bry06,wy08}. In Figure 
\ref{f:fit} it is interesting to include for comparison HD~69830 (HIP~40693), 
a $\sim2-4$~Gyr old, solar metallicity dwarf known to host a warm disk closer 
than $\sim1$~AU \citep{bei05} as well as three Neptune-mass planets within 
$0.6$~AU \citep{lo06}. Although HD~69830 has a $\teff$ somewhat cooler than the 
bulk of our stars \citep{sousa08}, it shows a clear excess in $W2$ and $W4$, 
while $W1$ and $W3$ are barely affected. A somewhat similar trend is also 
shown by HD~72905 (HIP~42438), a $\sim0.2$~Gyr old, solar-like star surrounded 
by hot dust \citep{bei06}\footnote{HD~69830 satisfies all photometric quality 
requirements listed in Section \ref{s:sample}, apart from missing 2MASS 
counterpart within $3$~arcsec. Notice though that this star has proper motion 
of $\sim1$~arcsec/year. HD~72905 was in our initial sample of Section 
\ref{s:sample}, but discarded because $\rm{ext}=1$. This flag implies that the 
profile fit of the photometry is not optimal (namely $\chi^2=3.6$ in $W3$), 
although it is still not associated with any 2MASS Extended Source and all 
other photometric quality requirements of Section \ref{s:sample} are 
satisfied.}. 

However, producing emission in $W2$ while keeping the other two contiguous 
bands essentially unaffected requires major fine tunings, thus rendering also 
the disk interpretation difficult. At first, interpreting all anomalous WISE 
colors as suspected and/or poor photometry seems difficult because of the 
various quality constraints imposed (Section \ref{s:sample}), although 
a number of W2 excesses have a saturated pixel fraction higher than usual 
(c.f.~Section \ref{s:sample}) and
more stringent cuts do alleviate the problem. \cite{kw} have conducted a 
throughout study of stars in the {\it Kepler} field using WISE photometry to 
identify 
disk candidates, and concluded in fact that spurious detections are less 
likely when using photometric measurements with source extraction $\chi^2$ 
smaller than $2,\,1.5,\,1.2,\,1.2$ in $W1-4$, respectively. The requirement 
$\rm{ext}=0$ we adopt is in fact  less stringent (Section \ref{s:sample}); the 
tighter $\chi^2$ constraints listed above are satisfied by $98\%$ ($W1$), 
$80\%$ ($W2$), $53\%$ ($W3$), $93\%$ ($W4$), of the stars used in Section 
\ref{s:main}, and by $100\%$ ($W1$), $50\%$ ($W2$), $50\%$ ($W3$), $95\%$ 
($W4$) of the stars in Table \ref{t:anom}. This suggests that any excess we 
see in $W4$ might be real, but it casts some doubts on $W2$. Adopting these 
constraints the number of outliers in Table \ref{t:anom} considerably reduces, 
although a number of them still remains: HIP~109110, objects with excess in 
$W2$ only (HIP~38228, HIP~88194 HIP~100963) and $W4$ only (HIP~38072 and 
HIP~66885). 

\begin{figure*}
\includegraphics[trim= 0mm -10mm 0mm -30mm, scale=0.6,keepaspectratio=true, angle=90]{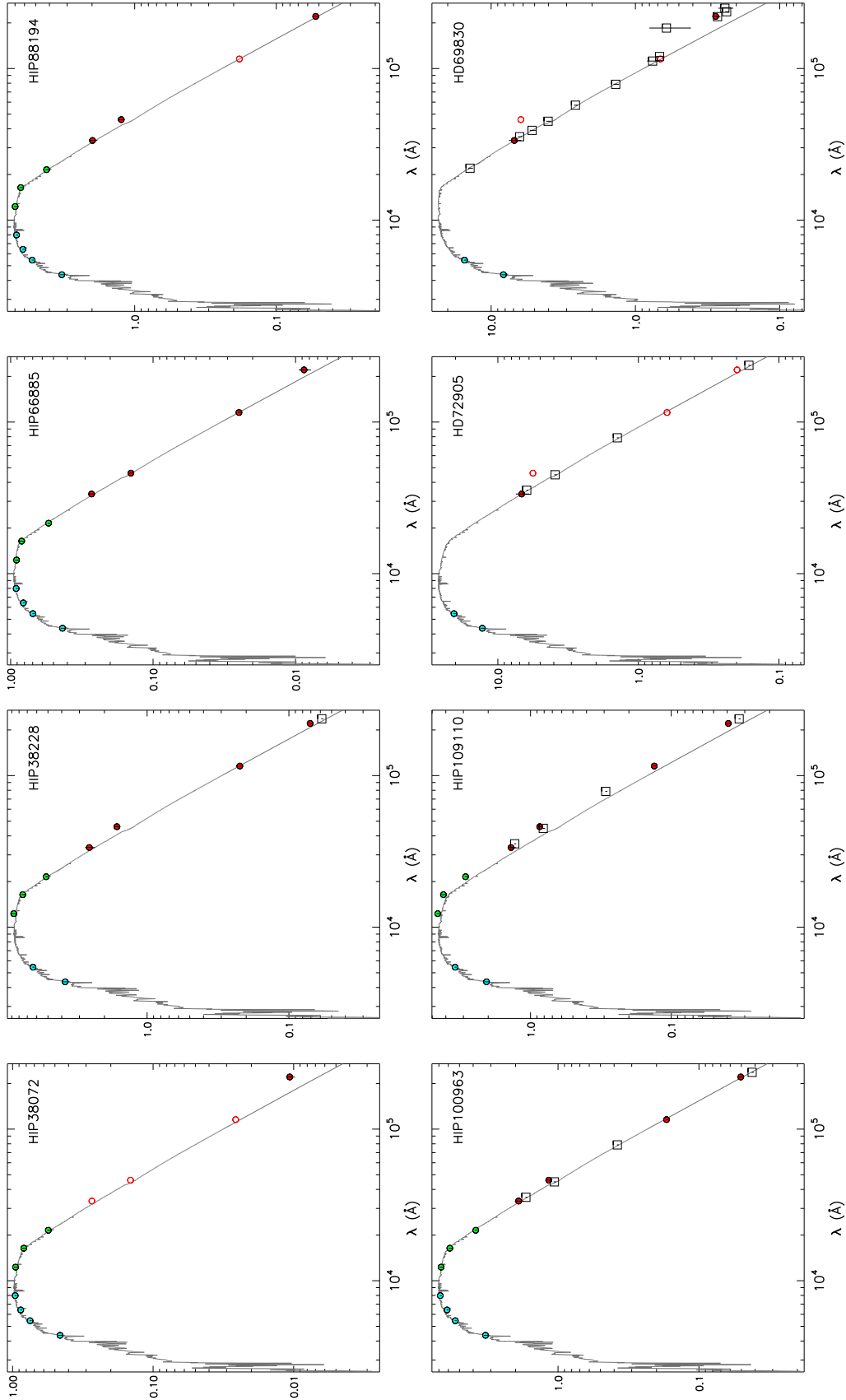}
\caption{Observed fluxes vs.~photospheric models \citep[from][]{ck04} for 
stars still having anomalous WISE colors after imposing more stringent 
requirements on the source extraction. Fluxes $-\,F_{\nu} \rm{(Jy)}\,-$ are 
derived from optical (cyan), 2MASS (green) and WISE (red) magnitudes as 
described in text: filled (open) symbols identify WISE photometric 
measurements that do (not) satisfy the requirements discussed in Section 
\ref{s:disc}, including the stricter $\chi^2$ extraction threshold. Open 
squares are independent flux measurements available in literature 
\citep{car08,pav09,bei11}.}
\label{f:sedfit}
\end{figure*}

For these six stars (plus HD~69830 and HD~72905), photospheric models tailored 
at the measured spectroscopic parameters are compared to fluxes derived from 
the adopted photometry (Figure \ref{f:sedfit}). The absolute calibration of 
those model fluxes is done by forcing them to return synthetic $V$ magnitudes 
that match the observed ones. We also tested on the full sample of stars in 
Section \ref{s:sample} that the mean difference in physical fluxes is $0.25\%$ 
and never exceeds $1.5\%$ if doing instead the absolute calibration using 
angular diameters obtained from the IRFM. These differences are essentially 
indistinguishable on the scale of the plot, and are taken into account when 
computing the flux uncertainty associated to each photometric measurement. 
Magnitudes in the $BV(RI)_\mathrm{C}\,JHK_s$ system are converted into fluxes 
using the same absolute calibration adopted in \cite{c10}, which has an 
intrinsic uncertainty of order $1-2\%$. Because of the aforementioned 
difference between using $V$ magnitudes or angular diameters, we have 
increased the global flux error to a conservative $3\%$. Similarly, for the 
$WISE$ system, we have adopted the absolute calibration and errors from 
\cite{jar11}, further increasing the latter by $1\%$, which also takes into 
account the possible zero-point offset discussed in Appendix \ref{s:app}. 
Observed magnitude errors are then added to the aforementioned uncertainties 
regarding the absolute flux of each band. As already expected from color 
indices, any difference with respect to photospheric models in $W2$ and $W4$ 
is significant, and it does not stem from uncertainties on the flux scale, nor 
in the observed magnitudes. The advantage of fitting photospheric models 
instead of using color indices is that we are now able to better quantify the 
observed anomalies.
The same comparison with synthetic spectra have also been done for all other 
stars in our sample not showing anomalous WISE colors, and indeed there are no 
mismatches between synthetic spectra and observations, thus validating the 
overall flux scale we adopt and also excluding major model deficiencies in 
those bands for solar-like stars \citep[but see][for possible model 
inaccuracies at cooler $\teff$]{kw}. 

As we already discussed, adding a cool companion does not modify the energy 
distribution in a way which is able to explain the observations, apart from 
HIP~109110, which longward of the $J$ band shows fluxes systematically higher 
than predicted. This likely indicates the presence of a cool companion (which 
we are able to fit with a model having $\teff \sim 4000$~K), an interpretation 
which is consistent with its suspected binarity \citep{f07} and with the 
linear trend observed in its radial velocity \citep{ni02}. This infrared 
excess is then further confirm by the {\it Spitzer-IRAC} and {\it -MIPS} 
measurements \citep{car08} shown in Figure \ref{f:sedfit} as open squares.

The $W2$ measurements for HD~69830 and HD~72905 do not pass the more 
stringent requirements we impose on the source extraction. The spurious 
nature of $W2$ photometry for these two stars is then confirmed by the absence 
of excess in other measurements at similar wavelengths \citep{car08,bei11}. 
Note though, for HD~69830 the {\it IRS}, {\it Spitzer-MIPS} and {\it IRAS} data 
\citep{bei11} confirm the excess we see in $W4$. Thus, we are inclined to 
regard any $W2$ excess among our stars as artificial, even if the source 
extraction is fine; this is further supported by the fact that in Figure 
\ref{f:sedfit} the $W2$ photometry of HIP~100963 does not agree with 
{\it Spitzer-IRAC} \citep{car08}. No additional measurements around 
$4.6 \mic$ exist for HIP~38228 and HIP~88194, but from the previous 
discussion, and because these excesses seem rather challenging to interpret 
when contiguos bands agree well with photospheric models, we conclude that 
also their nature is likely spurious. 

Finally, for HIP~38072 and HIP~66885 the deviation from photospheric models 
starts only in $W4$. Despite these two stars being the faintest among those in 
Table \ref{t:anom}, so far our adopted quality constraints have been enough to 
discard unreliable $W4$ measurements, and what we see could be indeed the 
signature of debris disks around these two stars. For HIP~38072 we have a 
measured flux of $10.7\pm0.7$~mJy and a photospheric prediction of 
$6.7\pm0.2$~mJy, thus resulting in an excess ratio of $1.6$ with a 
$5\sigma$ significance, while for HIP~66885 the measured flux is 
$8.7\pm0.8$~mJy versus a photospheric prediction of $7.0\pm0.2$~mJy, the 
excess ratio being $1.2$ at $2\sigma$. Using the absence of 
emission in $W3$ to constrain their temperature, we are able to easily fit 
these excesses with a black-body, but measurements at other wavelengths are 
clearly required to confirm or discard the presence of any disk. Should these 
two detections be confirmed, and using the variance of the the binomial 
distribution to derive a realistic error bar for such a low number statistic 
\citep[e.g.][]{be69} the occurrence rate of debris disk at $22\mic$ from our 
solar-like sample would thus be $3\pm2$\%, in good agreement with the 
$\sim4$\% estimated by \cite{tril08} at $24\mic$.

\begin{deluxetable}{cccccc}
\tablewidth{0pc}
\tabletypesize{\small}
\tablecaption{Stars having $V-W2 \ge 1.65$ or $V-W4 \ge 1.80$.
Bands that do not satisfy all of the quality constraints discussed in Section 
\ref{s:sample} are written in italics. The $\chi^2$ of the source extraction 
in each band is given in parenthesis.}
\tablehead{\colhead{HIP} & \colhead{$V$} & \colhead{$W1$} & \colhead{$W2$} & \colhead{$W3$} & \colhead{$W4$} }
\startdata
  8507          &  $8.899$ &  $7.185$         &   $7.217$        &   $7.230$        &   $7.109$  \\
 & & {\tiny $ (1.45)$}  & {\tiny $ (1.25)$}  & {\tiny $ (1.00)$}  & {\tiny $ (1.09)$}  \\
 11072         &  $5.190$ &  ${\it 3.607}$  &   $2.903$        &   $3.631$        &   $3.594$  \\
 & & {\tiny $ (0.28)$}  & {\tiny $ (2.09)$}  & {\tiny $ (2.12)$}  & {\tiny $ (1.04)$}  \\
 12186         &  $5.785$ &  $4.241$         &   $3.741$        &   $4.252$        &   $4.210$  \\
 &  & {\tiny $ (0.73)$}  & {\tiny $ (2.25)$}  & {\tiny $ (1.71)$}  & {\tiny $ (0.87)$}  \\
 15457          &  $4.836$ &  ${\it 3.334}$  &   ${\it 2.614}$ &   $3.334$        &   $3.243$  \\
 &  & {\tiny $ (0.38)$}  & {\tiny $ (1.21)$}  & {\tiny $ (1.66)$}  & {\tiny $ (0.96)$}  \\
 22263          &  $5.497$ &  $3.956$         &   $3.467$ 	      &   $3.981$        &   $3.912$  \\
 &  & {\tiny $ (0.37)$}  & {\tiny $ (2.04)$}  & {\tiny $ (1.37)$}  & {\tiny $ (0.93)$}  \\
 29525          &  $6.442$ &  $4.768$         &   $4.475$ 	      &   $4.857$        &   $4.800$  \\
 &  & {\tiny $ (0.33)$}  & {\tiny $ (1.70)$}  & {\tiny $ (0.76)$}  & {\tiny $ (0.91)$}  \\
 38072          &  $9.222$ &  ${\it 7.627}$  &   ${\it 7.678}$ &   ${\it 7.626}$ &   $7.227$  \\
 &  & {\tiny $ (1.57)$}  & {\tiny $ (0.90)$}  & {\tiny $ (1.26)$}  & {\tiny $ (1.21)$}  \\
 38228         &  $6.900$ &  $5.198$         &   $5.048$ 	      &   $5.293$        &   $5.170$  \\
 &  & {\tiny $ (0.42)$}  & {\tiny $ (1.30)$}  & {\tiny $ (0.79)$}  & {\tiny $ (1.17)$}  \\
 44713          &  $7.306$ &  ${\it 5.650}$  &   ${\it 5.629}$ &   ${\it 5.744}$ &   $5.717$  \\
 &  & {\tiny $ (0.49)$}  & {\tiny $ (1.26)$}  & {\tiny $ (1.14)$}  & {\tiny $ (0.96)$}  \\
 55459          &  $7.646$ &  $5.965$         &   $5.990$ 	      &   $6.045$        &   $5.994$  \\
 &  & {\tiny $ (0.63)$}  & {\tiny $ (1.53)$}  & {\tiny $ (0.92)$}  & {\tiny $ (0.89)$}  \\
 57291          &  $7.466$ &  $5.802$         &   $5.730$ 	      &   $5.831$        &   $5.694$  \\
 &  & {\tiny $ (1.05)$}  & {\tiny $ (1.25)$}  & {\tiny $ (1.35)$}  & {\tiny $ (1.16)$}  \\
 66885          &  $9.302$ &  $7.638$         &   $7.691$ 	      &   $7.665$        &   $7.445$  \\
 &  & {\tiny $ (1.05)$}  & {\tiny $ (0.88)$}  & {\tiny $ (1.13)$}  & {\tiny $ (0.92)$}  \\
 77052          &  $5.868$ &  $4.129$         &   $3.547$ 	      &   $4.166$        &   $4.143$  \\
 &  & {\tiny $ (0.65)$}  & {\tiny $ (2.02)$}  & {\tiny $ (1.73)$}  & {\tiny $ (0.83)$}  \\
 80337          &  $5.391$ &  $3.919$         &   ${\it 3.378}$ &   ${\it 3.902}$ &   $3.834$  \\
 &  & {\tiny $ (0.74)$}  & {\tiny $ (2.03)$}  & {\tiny $ (0.65)$}  & {\tiny $ (0.74)$}  \\
 85042          &  $6.287$ &  ${\it 4.676}$  &   ${\it 4.240}$ &   $4.708$        &   $4.657$  \\
 &  & {\tiny $ (0.89)$}  & {\tiny $ (2.14)$}  & {\tiny $ (0.85)$}  & {\tiny $ (0.82)$}  \\
 88194          &  $7.101$ &  $5.472$         &   $5.343$ 	      &   $5.506$        &   $5.492$  \\
 &  & {\tiny $ (0.94)$}  & {\tiny $ (1.32)$}  & {\tiny $ (1.74)$}  & {\tiny $ (1.15)$}  \\
 96895          &  $5.959$ &  ${\it 4.385}$  &   ${\it 4.004}$ &   ${\it 4.444}$ &   $4.412$  \\
 &  & {\tiny $ (0.47)$}  & {\tiny $ (1.66)$}  & {\tiny $ (0.78)$}  & {\tiny $ (0.92)$}  \\
100963         &  $7.089$ &  $5.517$         &   $5.416$ 	      &   $5.583$        &   $5.538$  \\
 &  & {\tiny $ (0.52)$}  & {\tiny $ (1.08)$}  & {\tiny $ (0.75)$}  & {\tiny $ (1.06)$}  \\
109110         &  $7.570$ &  $5.870$         &   $5.743$ 	      &   $5.857$        &   $5.810$  \\
 &  & {\tiny $ (1.22)$}  & {\tiny $ (1.01)$}  & {\tiny $ (1.18)$}  & {\tiny $ (0.92)$}  \\
113357         &  $5.467$ &  $3.881$         &   $3.361$ 	      &   $3.927$        &   $3.914$  \\
 &  & {\tiny $ (0.29)$}  & {\tiny $ (1.75)$}  & {\tiny $ (2.38)$}  & {\tiny $ (0.84)$}  \\
\enddata
\label{t:anom}
\end{deluxetable}

\section{CONCLUSIONS}

The uncertainty on the zero point of the effective temperature scale has been 
addressed using one of the most stringent photometric contraints available, 
i.e. the infrared colors of Sun, here determined in a model independent way 
via LDRs. Such analysis leads to an excellent agreement with the indices 
derived using the $\teff$ scale of \cite{c10}, thus confirming its accuracy. 
Effective temperatures have also been determined implementing the IRFM in the 
WISE system, thus validating the overall consistency between the 2MASS and 
WISE absolute calibration. This work also shows the importance of using solar 
twins for the absolute calibration of photometric quantities, something which 
is getting increasingly more important in the era of large all-sky photometric 
surveys.

However, while 2MASS magnitudes are very well suited for the purpose of 
determining $\teff$ --once stars affected by binarity are removed and/or using 
the full quality and flag information available in 2MASS to discard bad 
photometry-- this does not necessarily hold for the WISE data. Special 
attention on WISE's photometric quality flags and source extraction 
information must be paid, yet a number of excess emissions in $W2$ seem 
artificial. In this respect 2MASS magnitudes lie in a ``sweet spot'', enough 
in the red to sample the Planck tail, yet largely unaffected by contamination 
and/or flux excess, that being real or spurious.

At this stage it is still unclear if all WISE mid-infrared anomalies found are 
stemming from bad measurements or they are rather associate to real physical 
phenomena. Either cases being possible, those stars are clearly not 
representative of the Sun and have not been used to derive its colors. It is 
extremely difficult to interpret in a consistent manner objects showing intense 
excess in $W2$ only (and in fact, comparison with independent measurements 
confirms the spurious nature), while on the contrary it seems genuine for $W4$. 
This latter excess could be the signature of warm/hot debris disks, the best 
candidates from our sample being HIP~38072 and HIP~66885. Data at longer 
wavelengths are clearly needed: should HIP~38072 be confirmed, it would be the 
first solar-analog/twin ($T_{\rm{eff}}^{\rm{{\tiny spec}}}=5839$~K, 
$\logg=4.53$~dex, $\feh=0.06$~dex) found to host such a debris disk. This star 
is also relatively young ($2.4$~Gyr) and has $\sim 4$ times more lithium than 
the Sun \citep{bau10}, and it is included in our HARPS radial velocity 
monitoring, thus making it a potentially interesting target to gauge new 
insights into the planet--disk interaction. 

{\it Note added in proof.} The reason for the anomalously bright $W2$ 
magnitudes likely stems from problems in the profile fitting photometry for 
bright saturated stars, see 
http://wise2.ipac.caltech.edu/docs/release/allsky/expsup/\\
sec6$\_$3c.html

\acknowledgments
We thank an anonymous referee for a prompt report and G.~Kennedy for pointing 
out the WISE bias in the profile fitting photometry of bright stars.
I.R.'s work was performed under contract with the California Institute of 
Technology (Caltech) funded by NASA through the Sagan Fellowship Program. 
J.M. acknowledges support from FAPESP (2010/17510-3) and CNPq. This 
publication makes use of data products from the Two Micron All Sky Survey, 
which is a joint project of the University of Massachusetts and the Infrared 
Processing and Analysis Center/California Institute of Technology, funded by 
the National Aeronautics and Space Administration and the National Science 
Foundation. This publication makes use of data products from the Wide-field 
Infrared Survey Explorer, which is a joint project of the University of 
California, Los Angeles, and the Jet Propulsion Laboratory/California Institute 
of Technology, funded by the National Aeronautics and Space Administration.

\appendix

\section{The InfraRed Flux Method using WISE photometry}\label{s:app}

The availability of 2MASS and WISE photometry for most of the stars in our 
sample permits to run the IRFM using these two systems separately, and to 
check that consistent effective temperatures are derived. The implementation 
of the method is identical to that described in \cite{c06,c10}, i.e.~for each 
star the bolometric flux is recovered using all available broad-band 
optical/near-infrared/mid-infrared colors, while infrared fluxes needed to 
derive $\teff$ are now computed using 2MASS or WISE photometry, respectively.

\begin{figure*}
\includegraphics[scale=0.55]{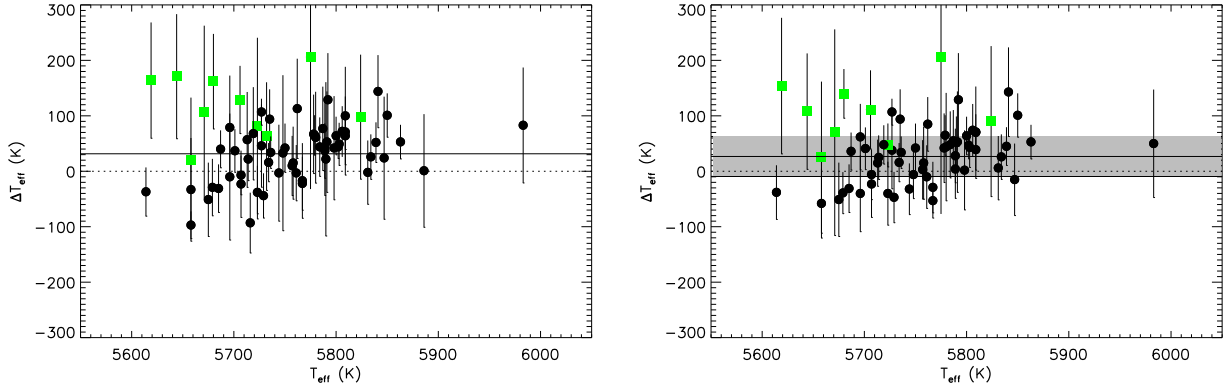}
\caption{$\Delta\teff(\rm{2MASS}-\rm{WISE})$ when implementing one or another 
system in the IRFM. Filled squares are stars having $V-W2$ and $V-W4$ redder 
than the thresholds discussed in Section \ref{s:sample}, and are not used to 
derive the mean difference (continuous line). Left panel: using in the IRFM at 
least two bands among $W1\,W2\,W3\,W4$. Right panel: when restricting to at 
least two bands among $W1\,W2\,W3$. Gray shaded area is the systematic offset 
($\pm 36$~K) allowed by the uncertainty in the WISE's absolute calibration.}
\label{f:2w}
\end{figure*}

For solar-type stars, the main driver in setting the zero-point of the $\teff$ 
scale is the absolute calibration of infrared bands. We have already discussed 
and tested that of 2MASS \citep[][and references therein]{c10}, thus focusing 
on the WISE system here. We adopt the $W1\,W2\,W3\,W4$ relative system response 
curves \footnote{http://wise2.ipac.caltech.edu/docs/release/allsky/expsup/sec4$\_$4h.html} and physical monochromatic fluxes $F_{\lambda}\rm{(iso)}$ from 
\cite{jar11}. The latter are built on the same absolute basis established for 
the {\it Spitzer Space Telescope}, ultimately constructed on the 
``Cohen-Walker-Witterborn '' framework \citep[][and references therein]{co03} 
and tied directly to the absolute mid-infrared calibrations by the 
{\it Midcourse Space Experiment} \citep[MSX,][]{price04}. These WISE fluxes 
have an expected overall systematic uncertainty of order $\sim 1.5\%$ and 
define the Vega zero-magnitude attributes upon which the effective temperatures 
we derive in this system directly depend. For each star we used the photometric 
constraints discussed in Section \ref{s:sample} and derived $\teff$ if at least 
two WISE bands were simultaneously available. $W4$ magnitudes have the 
largest photometric errors (Section \ref{s:sample}) thus showing the weakest 
correlation with $\teff$; indeed the scatter in the comparison with the 
effective temperatures obtained using 2MASS magnitudes reduces when $W4$ is 
not implemented in the IRFM (Figure \ref{f:2w}).

The maximum zero-point uncertainty in the effective temperatures derived using 
WISE photometry can be easily estimated by increasing or decreasing the adopted 
absolute calibration according to the error reported in \cite{jar11}. As 
expected, for the stellar parameter space covered here, the effect is 
essentially a constant offset of $\pm36\pm0.4$~K ($\sigma=3$~K). Notice that 
our reference 2MASS effective temperatures also have a zero-point uncertainty 
of order $20$~K \citep{c10}. The $25-30$~K mean difference when using 2MASS or 
WISE magnitudes in the IRFM is thus fully consistent within the uncertainties, 
and it would disappear if the WISE absolute calibration had been decreased by 
about $1\%$ (or conversely, the 2MASS absolute calibration increased by the 
same amount). Such a nice agreement is not entirely unexpected, since the 2MASS 
and WISE absolute calibration are built within the same 
``Cohen-Walker-Witterborn '' framework: nevertheless, the fact that the 2MASS 
absolute calibration has been independently verified using solar-twins 
confirms that the accuracy of the infrared absolute calibration extend also to 
the mid-infrared regime probed by WISE.

\clearpage

\setcounter{table}{0}
\LongTables
\tabletypesize{\footnotesize}
\begin{deluxetable}{lccrccccccl} 
\tablewidth{0pc}
\tablecaption{$(V-K_s)_\odot$ colors inferred from LDR Measurements. $N_\star$ is the number of stars used for the fit, $\sigma_{\rm{fit}}$ is the standard deviation of the fit minus data residual, and $\sigma_{\rm{ss}}$ is the standard deviation of the color inferred from the nine reflected Sun-ligh asteroid observations used for solar reference.}
\tablehead{\colhead{$\lambda_1$ (\AA)} & \colhead{species} & \colhead{$\lambda_2$ (\AA)} & \colhead{species} & \colhead{$N_\star$} & \colhead{$\sigma_\mathrm{fit}$} & \colhead{$(V-K_s)_\odot$} & \colhead{$\sigma_\mathrm{ss}$}}
\startdata
   5490.15 &  TiI & 5517.53 &  SiI & 85 & 0.038 & 1.555 & 0.011 \\
   5645.62 &  SiI & 5670.85 &   VI & 85 & 0.036 & 1.556 & 0.014 \\
   5650.71 &  FeI & 5670.85 &   VI & 85 & 0.036 & 1.558 & 0.009 \\
   5665.56 &  SiI & 5670.85 &   VI & 85 & 0.036 & 1.558 & 0.016 \\
   5665.56 &  SiI & 5703.59 &   VI & 85 & 0.031 & 1.567 & 0.016 \\
   5670.85 &   VI & 5701.11 &  SiI & 85 & 0.039 & 1.554 & 0.006 \\
   5690.43 &  SiI & 5703.59 &   VI & 85 & 0.037 & 1.552 & 0.017 \\
   5690.43 &  SiI & 5727.05 &   VI & 85 & 0.032 & 1.557 & 0.017 \\
   5701.11 &  SiI & 5703.59 &   VI & 85 & 0.036 & 1.557 & 0.006 \\
   5701.11 &  SiI & 5727.05 &   VI & 85 & 0.033 & 1.563 & 0.010 \\
   5701.11 &  SiI & 5727.65 &   VI & 84 & 0.040 & 1.559 & 0.004 \\
   5708.41 &  SiI & 5727.05 &   VI & 85 & 0.036 & 1.560 & 0.015 \\
   5727.05 &   VI & 5753.65 &  SiI & 85 & 0.030 & 1.573 & 0.006 \\
   5727.05 &   VI & 5772.15 &  SiI & 48 & 0.031 & 1.573 & 0.008 \\
   5727.65 &   VI & 5753.65 &  SiI & 84 & 0.039 & 1.561 & 0.007 \\
   5778.47 &  FeI & 5793.08 &  SiI & 67 & 0.040 & 1.557 & 0.006 \\
   5862.36 &  FeI & 5866.45 &  TiI & 78 & 0.037 & 1.559 & 0.009 \\
   6039.73 &   VI & 6046.00 &   SI & 85 & 0.037 & 1.550 & 0.012 \\
   6039.73 &   VI & 6052.68 &   SI & 85 & 0.037 & 1.563 & 0.008 \\
   6046.00 &   SI & 6064.63 &  TiI & 83 & 0.035 & 1.544 & 0.014 \\
   6046.00 &   SI & 6091.18 &  TiI & 85 & 0.038 & 1.547 & 0.013 \\
   6046.00 &   SI & 6093.14 &  CoI & 83 & 0.032 & 1.559 & 0.015 \\
   6052.68 &   SI & 6081.44 &   VI & 31 & 0.033 & 1.573 & 0.006 \\
   6052.68 &   SI & 6091.18 &  TiI & 85 & 0.037 & 1.558 & 0.009 \\
   6052.68 &   SI & 6093.14 &  CoI & 83 & 0.030 & 1.574 & 0.009 \\
   6055.99 &  FeI & 6085.27 &  FeI & 39 & 0.031 & 1.556 & 0.021 \\
   6064.63 &  TiI & 6091.92 &  SiI & 83 & 0.039 & 1.557 & 0.006 \\
   6078.50 &  FeI & 6085.27 &  FeI & 31 & 0.032 & 1.592 & 0.072 \\
   6085.27 &  FeI & 6086.29 &  NiI & 39 & 0.026 & 1.544 & 0.008 \\
   6085.27 &  FeI & 6155.14 &  SiI & 39 & 0.031 & 1.563 & 0.010 \\
   6089.57 &  FeI & 6090.21 &   VI & 83 & 0.031 & 1.561 & 0.005 \\
   6089.57 &  FeI & 6126.22 &  TiI & 83 & 0.037 & 1.553 & 0.006 \\
   6090.21 &   VI & 6091.92 &  SiI & 85 & 0.037 & 1.565 & 0.009 \\
   6090.21 &   VI & 6106.60 &  SiI & 85 & 0.041 & 1.566 & 0.003 \\
   6090.21 &   VI & 6125.03 &  SiI & 85 & 0.032 & 1.565 & 0.009 \\
   6090.21 &   VI & 6131.86 &  SiI & 78 & 0.036 & 1.562 & 0.011 \\
   6090.21 &   VI & 6155.14 &  SiI & 85 & 0.034 & 1.561 & 0.010 \\
   6091.92 &  SiI & 6111.65 &   VI & 84 & 0.038 & 1.556 & 0.008 \\
   6091.92 &  SiI & 6119.53 &   VI & 85 & 0.035 & 1.560 & 0.011 \\
   6091.92 &  SiI & 6126.22 &  TiI & 85 & 0.032 & 1.552 & 0.007 \\
   6091.92 &  SiI & 6128.99 &  NiI & 85 & 0.035 & 1.569 & 0.009 \\
   6106.60 &  SiI & 6111.65 &   VI & 84 & 0.037 & 1.566 & 0.014 \\
   6106.60 &  SiI & 6119.53 &   VI & 85 & 0.038 & 1.566 & 0.003 \\
   6106.60 &  SiI & 6126.22 &  TiI & 85 & 0.038 & 1.563 & 0.004 \\
   6106.60 &  SiI & 6135.36 &   VI & 84 & 0.037 & 1.571 & 0.012 \\
   6108.12 &  NiI & 6155.14 &  SiI & 85 & 0.031 & 1.572 & 0.007 \\
   6111.65 &   VI & 6131.86 &  SiI & 77 & 0.036 & 1.551 & 0.007 \\
   6119.53 &   VI & 6125.03 &  SiI & 85 & 0.031 & 1.559 & 0.007 \\
   6119.53 &   VI & 6131.86 &  SiI & 78 & 0.034 & 1.560 & 0.013 \\
   6091.92 &  SiI & 6111.65 &   VI & 84 & 0.038 & 1.556 & 0.008 \\
   6119.53 &   VI & 6142.49 &  SiI & 85 & 0.032 & 1.560 & 0.010 \\
   6119.53 &   VI & 6145.02 &  SiI & 85 & 0.032 & 1.559 & 0.007 \\
   6119.53 &   VI & 6155.14 &  SiI & 85 & 0.037 & 1.556 & 0.009 \\
   6125.03 &  SiI & 6126.22 &  TiI & 85 & 0.029 & 1.551 & 0.004 \\
   6125.03 &  SiI & 6128.99 &  NiI & 85 & 0.034 & 1.575 & 0.007 \\
   6126.22 &  TiI & 6131.86 &  SiI & 78 & 0.031 & 1.553 & 0.008 \\
   6126.22 &  TiI & 6142.49 &  SiI & 85 & 0.032 & 1.554 & 0.006 \\
   6126.22 &  TiI & 6145.02 &  SiI & 85 & 0.033 & 1.555 & 0.008 \\
   6126.22 &  TiI & 6155.14 &  SiI & 85 & 0.037 & 1.554 & 0.009 \\
   6128.99 &  NiI & 6131.86 &  SiI & 78 & 0.037 & 1.568 & 0.015 \\
   6128.99 &  NiI & 6142.49 &  SiI & 85 & 0.031 & 1.576 & 0.013 \\
   6128.99 &  NiI & 6145.02 &  SiI & 85 & 0.032 & 1.575 & 0.010 \\
   6151.62 &  FeI & 6155.14 &  SiI & 85 & 0.034 & 1.566 & 0.004 \\
   6155.14 &  SiI & 6180.22 &  FeI & 78 & 0.038 & 1.559 & 0.003 \\
   6155.14 &  SiI & 6199.19 &   VI & 71 & 0.039 & 1.555 & 0.008 \\
   6175.42 &  NiI & 6224.51 &   VI & 78 & 0.038 & 1.561 & 0.006 \\
   6176.81 &  NiI & 6224.51 &   VI & 78 & 0.038 & 1.561 & 0.006 \\
   6186.74 &  NiI & 6224.51 &   VI & 29 & 0.033 & 1.572 & 0.004 \\
   6199.19 &   VI & 6230.09 &  NiI & 71 & 0.038 & 1.549 & 0.006 \\
   6204.64 &  NiI & 6224.51 &   VI & 78 & 0.038 & 1.556 & 0.006 \\
   6204.64 &  NiI & 6243.11 &   VI & 85 & 0.028 & 1.558 & 0.006 \\
   6204.64 &  NiI & 6251.82 &   VI & 85 & 0.035 & 1.542 & 0.015 \\
   6215.15 &  FeI & 6243.11 &   VI & 85 & 0.037 & 1.558 & 0.010 \\
   6215.15 &  FeI & 6251.82 &   VI & 85 & 0.040 & 1.558 & 0.005 \\
   6223.99 &  NiI & 6224.51 &   VI & 78 & 0.038 & 1.559 & 0.004 \\
   6223.99 &  NiI & 6243.11 &   VI & 84 & 0.030 & 1.554 & 0.012 \\
   6223.99 &  NiI & 6251.82 &   VI & 84 & 0.039 & 1.549 & 0.013 \\
   6224.51 &   VI & 6230.09 &  NiI & 78 & 0.037 & 1.560 & 0.007 \\
   6230.09 &  NiI & 6243.11 &   VI & 85 & 0.033 & 1.564 & 0.008 \\
   6230.09 &  NiI & 6251.82 &   VI & 85 & 0.035 & 1.550 & 0.011 \\
   6237.33 &  SiI & 6240.66 &  FeI & 85 & 0.036 & 1.567 & 0.010 \\
   6237.33 &  SiI & 6243.11 &   VI & 85 & 0.034 & 1.559 & 0.012 \\
   6237.33 &  SiI & 6261.10 &  TiI & 85 & 0.031 & 1.555 & 0.012 \\
   6240.66 &  FeI & 6243.81 &  SiI & 85 & 0.039 & 1.565 & 0.022 \\
   6240.66 &  FeI & 6244.48 &  SiI & 85 & 0.039 & 1.566 & 0.019 \\
   6243.11 &   VI & 6243.81 &  SiI & 85 & 0.031 & 1.558 & 0.006 \\
   6243.11 &   VI & 6244.48 &  SiI & 85 & 0.031 & 1.559 & 0.010 \\
   6243.81 &  SiI & 6251.82 &   VI & 85 & 0.038 & 1.552 & 0.008 \\
   6243.81 &  SiI & 6261.10 &  TiI & 85 & 0.033 & 1.550 & 0.018 \\
   6244.48 &  SiI & 6261.10 &  TiI & 85 & 0.031 & 1.551 & 0.017 \\
   6327.60 &  NiI & 6414.99 &  SiI & 32 & 0.035 & 1.577 & 0.004 \\
   6330.13 &  CrI & 6330.86 &  FeI & 85 & 0.030 & 1.560 & 0.014 \\
   6330.13 &  CrI & 6414.99 &  SiI & 32 & 0.038 & 1.577 & 0.003 \\
   6392.55 &  FeI & 6414.99 &  SiI & 32 & 0.039 & 1.573 & 0.007 \\
   6414.99 &  SiI & 6498.95 &  FeI & 32 & 0.036 & 1.576 & 0.005 \\
   6703.57 &  FeI & 6721.85 &  SiI & 85 & 0.036 & 1.565 & 0.007 \\
   6710.31 &  FeI & 6721.85 &  SiI & 85 & 0.039 & 1.555 & 0.004 \\
   6710.31 &  FeI & 6748.84 &   SI & 84 & 0.037 & 1.557 & 0.012 \\
   6710.31 &  FeI & 6757.17 &   SI & 85 & 0.034 & 1.558 & 0.014 \\
   6746.96 &  FeI & 6757.17 &   SI & 65 & 0.030 & 1.555 & 0.010 \\

\enddata
\label{t:ldr}
\end{deluxetable}

\end{document}